\documentclass[%
 reprint,
showpacs,preprintnumbers,
 amsmath,amssymb,
aps,
]{revtex4-1}

\usepackage{graphicx}% Include figure files
\usepackage{xcolor}
\usepackage{dcolumn}% Align table columns on decimal point
\usepackage{bm}% bold math
\usepackage{hyperref}% add hypertext capabilities
\hypersetup{
    colorlinks=true,
    linkcolor=blue,
    citecolor=blue,     
    urlcolor=blue,
}
\usepackage{amsmath}
\mathchardef\mhyphen="2D

\allowdisplaybreaks

\begin{document}
%\def\arraystretch{2}\tabcolsep=5pt
%\preprint{APS/123-QED}

\title{Landau theory of metal-insulator transition in VO$_\text{2}$ doped with metal ions}% Force line breaks with \\

\author{Yin Shi}
 \email{yxs187@psu.edu}
\author{Long-Qing Chen}%
 \email{lqc3@psu.edu}
\affiliation{%
Department of Materials Sciences and Engineering, Pennsylvania State University, University Park, PA 16802, USA
}%

%\affiliation{
%\affiliation{
%\affiliation{%

\date{\today}% It is always \today, today,

\begin{abstract}
Metal-ion doping can effectively regulate the metal-insulator transition temperature in $\mathrm{VO}_2$.
Experiments found that the pentavalent and hexavalent ion doping dramatically reduces the transition temperature while the trivalent ion doping increases the transition temperature and induces intermediate phases.
Based on the phase-field model of the metal-insulator transition in $\mathrm{VO}_2$ we developed previously, we formulate a Landau potential of the metal-ion-doped $\mathrm{VO}_2$ taking account of the effects of doping on the electron correlation and lattice structure.
The effect of metal-ion doping on the lattice structure is accounted for in a phenomenological way.
Using the Landau potential, we calculate the temperature-dopant-concentration phase diagrams of $\mathrm{VO}_2$ doped with various metal ions consistent with the experiments and provide explanation to the different behaviors of different metal-ion doping.
The phenomenological theory can provide estimations of phase diagrams of $\mathrm{VO}_2$ doped with other metal ions.

%\pacs{64.60.Ej, 71.27.+a, 71.30.+h}

%\begin{description}
%\end{description}
\end{abstract}

\maketitle

%\tableofcontents

\section{Introduction}
Vanadium dioxide ($\mathrm{VO}_2$) exhibits a metal-insulator transition (MIT) at a temperature $T_c=338~\mathrm{K}$~\cite{Morin59Oxides,Park13Measurement}, which features novel device applications such as sensors, Mott field-effect transistors and memristors~\cite{Liao17Ultrafast,Nakano12Collective,Son11Excellent,Driscoll09Phase,Yang11Oxide}.
For temperatures above $T_c$, it is a metal with a rutile (R) structure, while for temperatures below $T_c$, it is an insulator with a monoclinic (M1) structure.
The application of the MIT may require the regulation of the transition temperature $T_c$, e.g., to around the room temperature.
In addition to applying stress~\cite{Muraoka02Metal,Fan14Strain} and controlling the microstructure and defects~\cite{Sahana02Microstructure,Luo13Optimization,Whittaker09Depressed}, an effective and convenient route to modulating $T_c$ is to dope $\mathrm{VO}_2$ with metal ions~\cite{Tan12Unraveling,Piccirillo07Nb}.
One can dramatically reduce $T_c$ by doping $\mathrm{VO}_2$ with larger higher-valence ions such as $\mathrm{W}^{6+}$, $\mathrm{Mo}^{5+}$ and $\mathrm{Nb}^{5+}$ (compared to $\mathrm{V}^{4+}$), or increase $T_c$ by doping $\mathrm{VO}_2$ with lower-valence ions such as $\mathrm{Cr}^{3+}$, $\mathrm{Al}^{3+}$ and $\mathrm{Ca}^{3+}$ (compared to $\mathrm{V}^{4+}$)~\cite{Piccirillo07Nb,Wu14Depressed,Patridge12Elucidating,Brown13Electrical,Wu14A,Beteille98Optical,Macchesney69Growth}.

Metal-ion doping introduces additional free carriers and changes in the lattice structure in the parent $\mathrm{VO}_2$, which are responsible for the regulation of $T_c$.
In the case of $\mathrm{W}^{6+}$ doping, W atoms substitute V atoms and transfer electrons to $\mathrm{V}^{4+}$ ions to form $\mathrm{V}^{3+}$ ions~\cite{Tang85Local}.
It disrupts the bonds between the Peierls-paired $\mathrm{V}^{4+}$ ions; instead, the $\mathrm{W}^{6+}$--$\mathrm{V}^{3+}$ and $\mathrm{V}^{4+}$--$\mathrm{V}^{3+}$ bonds develop, making the local structure around $\mathrm{W}^{6+}$ ions tetragonal~\cite{Booth09Anisotropic,Tan12Unraveling}.
$\mathrm{W}^{6+}$ doping also induces a significant expansion in the $[110]_\mathrm{R}$ and $[1\bar{1}0]_\mathrm{R}$ directions~\cite{Booth09Anisotropic}.
This change in the lattice structure destabilizes the M1 structure and thus lowers the thermal energy barrier for the transition from the M1 structure to the R structure~\cite{Whittaker11Distinctive,Netsianda08The,Booth09Anisotropic}.
This disruption of the Peierls pairing upon electron doping essentially rises from the electron-lattice coupling.
It is expected, since the addition of electrons suppresses the stability of the Peierls bonding.
In an equally-spaced-atom chain, with the number of electrons per atom increasing, fewer atoms have their interatomic bonds shrunk to form a unit cell with two electrons through the Peierls transition, and finally the bond shrinking becomes totally unfavorable for two electrons per atom~\cite{Kagoshima81Peierls}.

Besides its influence on the electronic structure through the electron-lattice coupling, the additional free charges introduced by doping can affect the electronic structure of the  interacting electrons in $\mathrm{VO}_2$ via modifying the electron-electron interaction, which however does not get much attention from researchers.
Indeed, free charges screen the electron-electron repulsion, thereby weakening the Mott instability~\cite{Wegkamp14Instantaneous,Kim04Mechanism,Stefanovich00Electrical,Nakano12Collective,Mazza16Field}.

On the other hand, metal-ion doping can directly distort the lattice structure due to the different size of the dopant ion from that of the $\mathrm{V}^{4+}$ ion.
The radius of the $\mathrm{W}^{6+}$ ion is larger than that of the $\mathrm{V}^{4+}$ ion, and indeed it was found that $\mathrm{W}^{6+}$ doping induces an increase in cationic spacing in the lattice~\cite{Booth09Anisotropic}.
The trend of the change in $T_c$ can be correlated with the relative size of the dopant ion compared to that of the $\mathrm{V}^{4+}$ ion~\cite{Macchesney69Growth}.

The detailed mechanism of the increase of $T_c$ due to trivalent-ion ($\mathrm{Cr}^{3+}$, $\mathrm{Al}^{3+}$ and $\mathrm{Ca}^{3+}$) doping is difficult to elucidate.
Unlike in the higher-valence-ion doping where only the M1 and the R phases are involved in the MIT, doping $\mathrm{VO}_2$ with lower-valence ions ($\mathrm{Cr}^{3+}$ or $\mathrm{Al}^{3+}$) induces an intermediate insulating monoclinic (M2) phase and an intermediate insulating triclinic (T) phase besides the M1 and the R phases~\cite{Villeneuve73Contribution,Marezio72Structural,Pouget76Metal,Strelcov12Doping,Ghedira77Structural}.

The entanglement of effects of the charge doping and the lattice distortion induced by metal-ion doping complicates the understanding of the roles of the two effects in the regulation of $T_c$.
Doping $\mathrm{VO}_2$ with the tetravalent ion $\mathrm{Ti}^{4+}$ can rule out the effect of the charge doping.
The local structure around $\mathrm{Ti}^{4+}$ ions in the M1 phase is found to be that of the anatase and the $\mathrm{VO}_6$ octahedra is subtly distorted by $\mathrm{Ti}^{4+}$ doping~\cite{Wu15Decoupling}.
$\mathrm{Ti}^{4+}$ doping shows limited ability to regulate $T_c$~\cite{Wu15Decoupling}, which indicates that the charge doping is more effective than the lattice distortion in regulating $T_c$~\cite{Wu15Decoupling}.

To understand these phenomena, we use the phenomenological theory to study the effect of metal-ion doping on the MIT in $\mathrm{VO}_2$.
The phase-field model we formulated previously can account for the effect of the charge doping on the electron correlation~\cite{Shi17Ginzburg,Shi18Phase,Shi19Current}.
To address the full effect of the metal-ion doping, we take into account the effect of doping on the lattice structure in the phenomenological way.
With this, we calculate the temperature-dopant-concentration phase diagrams of $\mathrm{VO}_2$ doped with various metal ions consistent with experiments and provide insights into the distinct behaviors of the pentavalent (or hexavalent) ion doping and the trivalent ion doping.

\section{Effect of charge doping on electron correlation}
In previous works~\cite{Shi17Ginzburg,Shi18Phase,Shi19Current}, we formulated a phase-field model of the MIT in $\mathrm{VO}_2$.
In this work, we utilize this model and only consider the homogeneous case.
For a homogeneous system without the presence of electric fields, the thermodynamics of the MIT is described by a Landau-type potential-energy density,
\begin{align}
F(T,\Phi;\{\eta_i\},\{\mu_i\},n,p) =& F_0(T;\{\eta_i\},\{\mu_i\}) \notag  \\
& + F_{e \mhyphen h}(T;\{\mu_i\},n,p),
\label{eq:F}
\end{align}
which consists of a contribution from the intrinsic $\mathrm{VO}_2$, $F_0$, and a contribution from additional free carriers, $F_{e \mhyphen h}$.
Here $T$ is the temperature, $\eta_i~(i=1,2,3,4)$ are the structural order parameters, $\mu_i~(i=1,2,3,4)$ are the electronic order parameters, and $n$ and $p$ are the free-electron and free-hole densities, respectively.
A finite $\eta_i$ indicates the dimerization of the neighboring V atoms, and a finite $\mu_i$ indicates the formation of the dynamical singlet situated on the neighboring V sites and consequently the opening of the energy gap~\cite{Biermann05Dynamical,Zheng15Computation,Brito16Metal}. 
The order parameters of the different phases are: $\eta_1=\eta_3\neq 0, \eta_2=\eta_4=0, \mu_1=\mu_3\neq 0, \mu_2=\mu_4=0, \eta_1\mu_1<0, \eta_3\mu_3<0$ (and other symmetry-related values) for the M1 phase, $\eta_1\neq 0, \eta_2=\eta_3=\eta_4=0, \mu_1\neq 0, \mu_2=\mu_3=\mu_4=0, \eta_1\mu_1<0$ (and other symmetry-related values) for the M2 phase, and $\eta_i=0,\mu_i=0,i=1,2,3,4$ for the R phase~\cite{Shi17Ginzburg}.

$F_0$ is a Landau expansion on the order parameters~\cite{Shi17Ginzburg},
\begin{align}
F_0=& \frac{a(T-T_0)}{2T_c}\eta_i\eta_i + \frac{b_{ij}}{4}\eta_i^2\eta_j^2 + \frac{c_{ij}}{6}\eta_i^2\eta_j^4 \notag \\
&+ \frac{A(T-T_0')}{2T_c}\mu_i\mu_i + \frac{B_{ij}}{4}\mu_i^2\mu_j^2 + \frac{C_{ij}}{6}\mu_i^2\mu_j^4 \notag \\
&+ h\eta_i\mu_i - \frac{p_{ijkl}}{2}\eta_i\eta_j\mu_k\mu_l + \frac{q_{ijkl}}{2}\eta_i\eta_j\eta_k\mu_l,
\end{align}
where $T_0$ and $T_0'$ are the ``Curie-Weiss temperatures'' of the structural and the electronic order parameters, respectively, and $a$, $b_{ij}$, $c_{ij}$, $A$, $B_{ij}$, $C_{ij}$, $h$, $p_{ijkl}$ and $q_{ijkl}$ are constants satisfying certain symmetry relations~\cite{Shi17Ginzburg}.
The Einstein summation convention has been used.
$F_{e \mhyphen h}$ is
\begin{align}
F_{e \mhyphen h}= & k_BT\left[ \int_0^n F_{1/2}^{-1}\left(\frac{n'}{N_c}\right)dn'+\int_0^p F_{1/2}^{-1}\left(\frac{p'}{N_v}\right)dp' \right]  \notag  \\
& + \frac{E_g}{2}(n+p) - F_\mathrm{in}(T;\{\mu_i\}),
\label{eq:G}
\end{align}
where $F_{1/2}^{-1}$ represents the inverse function of the Fermi integral $F_{1/2}(x)=(2/\sqrt{\pi})\int_0^\infty\sqrt{\epsilon}[1+\exp(\epsilon-x)]^{-1}d\epsilon$, $k_B$ is the Boltzmann constant, and $N_c$ and $N_v$ are the effective densities of states of the conduction and valence bands, respectively.
$E_g$ is the energy gap directly related to the electronic order parameters~\cite{Biermann05Dynamical,Zheng15Computation,Brito16Metal} $E_g(\{\mu_i\})\approx 2U^2\mu_0^2\sum_i\mu_i^2/k_B T_c$, where $U$ is the on-site Coulomb repulsion and $\mu_0$ is a dimensionless parameter~\cite{Shi17Ginzburg,Shi18Phase}. 
$F_\mathrm{in}$ is the equilibrium intrinsic free energy of the electrons and holes, and thus $F_{e \mhyphen h}$ vanishes at equilibrium.
It satisfies $\partial F_\mathrm{in}/\partial \mu_i = n_\mathrm{in}d E_g/d \mu_i$, where $n_\mathrm{in}=N_cF_{1/2}[(\xi_\mathrm{in}-E_g/2)/k_BT]$ is the intrinsic carrier density ($\xi_\mathrm{in}$ is the equilibrium intrinsic chemical potential of free electrons)~\cite{Shi19Current}.

The equilibrium state is determined by the minimum of $F$ with respect to the order parameters
\begin{equation}
\frac{\partial F}{\partial \eta_i} = 0,~\frac{\partial F}{\partial \mu_i} = 0~(i=1,2,3,4),
\label{eq:minF}
\end{equation}
and by the equilibrium quasi-chemical potentials of free electrons and holes
\begin{equation}
\xi_e = \frac{\partial F}{\partial n} = \xi_0,~\xi_h = \frac{\partial F}{\partial p} = -\xi_0,
\label{eq:eqchempot}
\end{equation}
where $\xi_0$ is the equilibrium chemical potential of free electrons.
The solution to Eq.~(\ref{eq:eqchempot}) is just $n=N_c F_{1/2}[(\xi_0-E_g/2)/k_BT]$ and $p=N_v F_{1/2}[(-\xi_0-E_g/2)/k_BT]$.

\begin{figure*}[t!]
\centering
\includegraphics[width=0.8\textwidth]{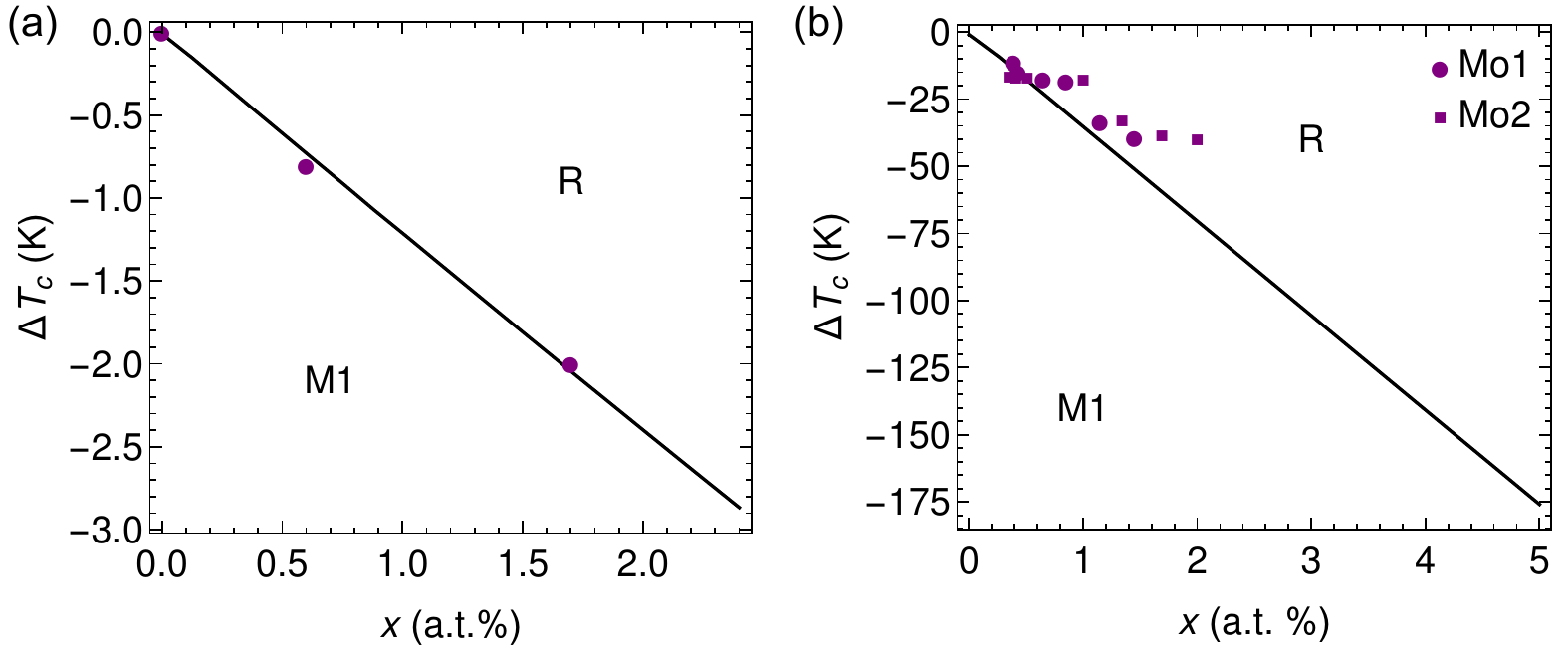}
\caption{Temperature versus doping concentration phase diagrams of (a) $\mathrm{V}_{1-x}\mathrm{Ti}_x\mathrm{O}_2$ and (b) $\mathrm{V}_{1-x}\mathrm{Mo}_x\mathrm{O}_2$. $\Delta T_c$ is the difference of the MIT temperature from $T_c=338~\mathrm{K}$. In (a), the purple dots are the experimentally measured R-M1 phase boundary~\cite{Wu15Decoupling} and the black line is the calculated R-M1 phase boundary with $\Theta_1=16 k_B T_c$. In (b), the purple markers are the experimentally measured R-M1 phase boundary: Mo1~\cite{Patridge12Elucidating} and Mo2~\cite{Whittaker11Microscopic}. The black line is the calculated R-M1 phase boundary with $\Theta_1=16 k_B T_c$ and $\Theta_2 = 20 k_B T_c$.}
\label{fig:TiMo}
\centering
\end{figure*}

The effect of charge doping on the electron correlation is reflected by the influence of $n$ and $p$ on the electronic order parameters.
If $\mathrm{VO}_2$ is electron-doped with an electron density $N_d$ and also hole-doped with a hole density $N_a$, $\xi_0$ is determined from the charge neutrality condition $n+N_a=p+N_d$ such that $n \approx N_d-N_a \gg p \approx n_\mathrm{in}^2/ (N_d-N_a)$ for $N_d-N_a \gg n_\mathrm{in}$ or $p \approx N_a-N_d \gg n \approx n_\mathrm{in}^2/ (N_a-N_d)$ for $N_a-N_d \gg n_\mathrm{in}$~\cite{Moll64Physics}.
Then from Eq.~(\ref{eq:minF}) we have
\begin{align}
    \frac{\partial F}{\partial \mu_i} &= \frac{\partial F_0}{\partial \mu_i} + \frac{dE_g}{d\mu_i}\left(\frac{n+p}{2} - n_\mathrm{in}\right) \notag \\
    &\approx \frac{\partial F_0}{\partial \mu_i} + \frac{2U^2\mu_0^2|N_a-N_d|\mu_i}{k_BT_c}~(i=1,2,3,4).
\end{align}
The second term on the right-hand side of the equation renormalizes down $T_0'$,
\begin{equation}
T_0' \rightarrow T_0' - \frac{2U^2\mu_0^2|N_a-N_d|}{k_BA},
\label{eq:renormT0doped}
\end{equation}
indicating that the effect of charge doping on the electron correlation is to assist the transition from an insulator to a metal.

\section{Effects of doping on lattice structure}
The effects of doping on the lattice structure are at least two-fold.
The radius of the dopant ion is different from that of the $\mathrm{V}^{4+}$ ion, which may induce expansion or shrinkage depending on the relative size of the dopant ion with respect to the size of the $\mathrm{V}^{4+}$ ion.
This can be termed as the volume effect.
On the other hand, the additional free charges introduced by the dopants may have an impact on the Peierls instability, which we term as the Peierls effect here.
The volume effect may be characterized by an energy of the coupling between the relative volume of the dopant ion and the structural order parameters $\eta_i$.
We only consider the case of dilute doping so that only the contribution on the lowest order of the dopant concentration is important to the coupling energy.
Since the volume is a scalar, the coupling energy on the lowest order of the dopant-induced relative change in volume and on the lowest order of $\eta_i$ is
\begin{equation}
    F_1 = \frac{1}{2}\Theta_1 x\frac{R_d^3-R_{\mathrm{V}^{4+}}^3}{R_{\mathrm{V}^{4+}}^3}(\eta_1^2+\eta_2^2+\eta_3^2+\eta_4^2),
\end{equation}
where $\Theta_1$ is a coupling constant, $x$ is the atomic fraction of the dopant with respect to V atom, $R_d$ is the radius of the dopant ion, and $R_{\mathrm{V}^{4+}}$ is the radius of the $\mathrm{V}^{4+}$ ion.

The Peierls effect is rather obscure to describe.
To describe it, we first look into the Peierls transition.
An equally-spaced-atom chain with a lattice constant $b$ and $n_0$ electrons per atom is unstable for temperatures below some critical value.
Periodic lattice distortion with a wavelength $2b/n_0$ develops, in which every neighboring $2/n_0$ atoms get closer to form a unit cell with $2$ electrons~\cite{Kagoshima81Peierls}.
Therefore, an increase (decrease) in $n_0$ tends to weaken (strengthen) the Peierls distortion.
We may account for this tendency by a coupling energy
\begin{equation}
    F_2 = \frac{1}{2}\Theta_2 x(v_d - v_{\mathrm{V}^{4+}})(\eta_1^2+\eta_2^2+\eta_3^2+\eta_4^2),
\end{equation}
where $\Theta_2$ is a coupling constant, $v_d$ is the valence of the dopant ion, and $v_{\mathrm{V}^{4+}}=4$ is the valence of the vanadium in $\mathrm{VO}_2$.
Again, we only considered the coupling energy on the lowest order of the density of excess electrons and on the lowest order of $\eta_i$.

$F_1$ and $F_2$ added to $F$ in Eq.~(\ref{eq:F}) renormalize $T_0$,
\begin{equation}
    T_0 \rightarrow T_0 - \frac{T_c[\Theta_1(R_d^3-R_{\mathrm{V}^{4+}}^3)/R_{\mathrm{V}^{4+}}^3+\Theta_2(v_d - v_{\mathrm{V}^{4+}})]x}{a}.   \label{eq:renormT0struct}
\end{equation}

The effect of doping on the lattice structure is more subtle than what we described by Eq.~(\ref{eq:renormT0struct}), e.g., dopant ions can change the local lattice structure~\cite{Booth09Anisotropic,Wu15Decoupling} which may correspond to the emergence of inhomogeneous eigenstrain distribution near the dopant ions.
However, the idea here is to characterize the doping effects by simple characters (the radius and the valence) of the dopant ion.
We treated the doped system as homogeneous in mesoscale, which corresponds to an average of properties over microscale.
This reconciles with the coarse-grained nature of the Landau theory.

\begin{figure*}[t!]
\centering
\includegraphics[width=0.8\textwidth]{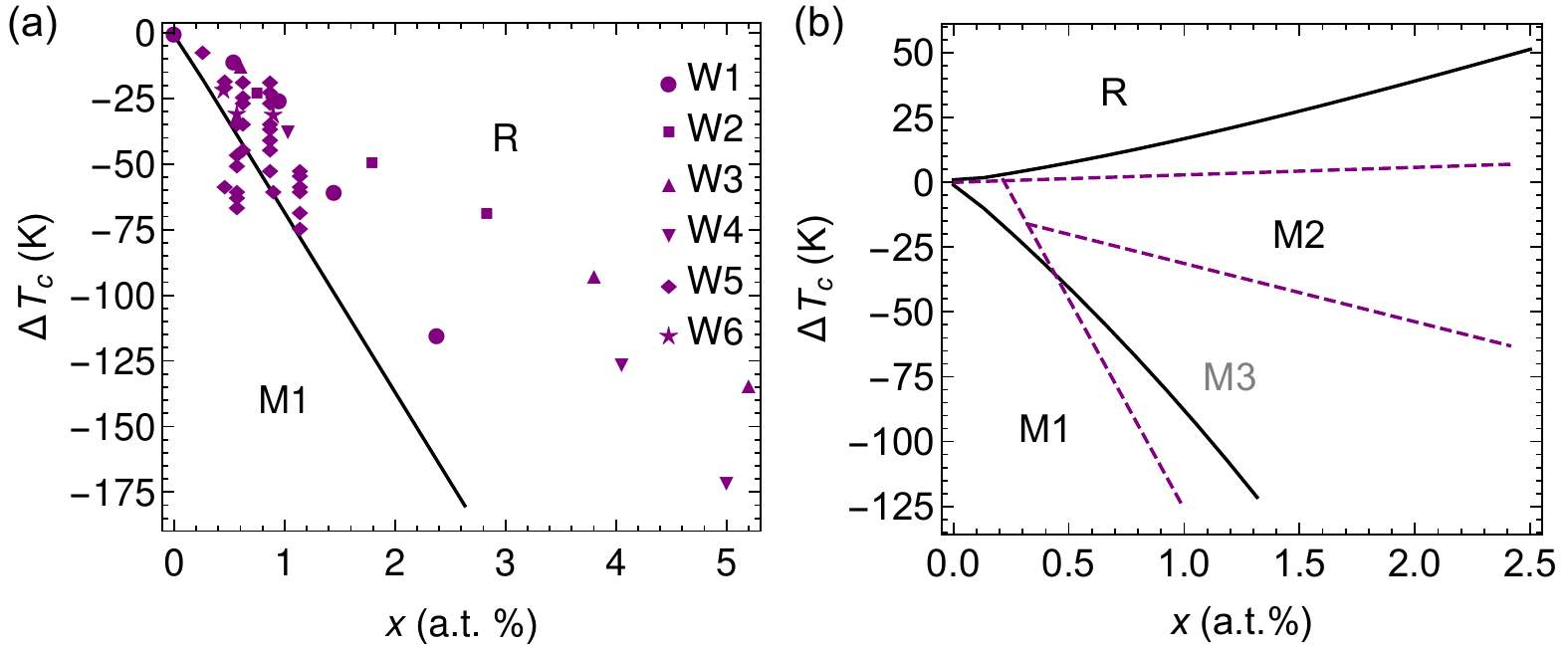}
\caption{Temperature versus doping concentration phase diagrams of (a) $\mathrm{V}_{1-x}\mathrm{W}_x\mathrm{O}_2$ and (b) $\mathrm{V}_{1-x}\mathrm{Cr}_x\mathrm{O}_2$. $\Delta T_c$ is the difference of the MIT temperature from $T_c=338~\mathrm{K}$. In (a), the purple markers are the experimentally measured R-M1 phase boundary: W1~\cite{Tan12Unraveling}, W2~\cite{Shi07Preparation}, W3~\cite{Horlin72Electrical}, W4~\cite{Shibuya10Metal}, W5~\cite{Wu11Temperature}, and W6~\cite{Whittaker11Distinctive}. The black line is the calculated R-M1 phase boundary. In (b), the purple dashed lines are the experimentally measured phase boundaries~\cite{Marezio72Structural}, and the black lines are the calculated phase boundaries. The experiment found a new monoclinic (M3) phase between the M1 and the M2 phases~\cite{Marezio72Structural} (see text).}
\label{fig:WCr}
\centering
\end{figure*}

\section{Phase diagrams of $\mathrm{VO}_2$ doped with various dopants}
Let us consider a doped system $\mathrm{V}_{1-x}\mathrm{M}_x\mathrm{O}_2$, that is, $\mathrm{VO}_2$ doped with metal-ion $\mathrm{M}^{v_d+}$ of $x$ atomic fraction.
We use Eqs.~(\ref{eq:renormT0doped},\ref{eq:renormT0struct}) to simulate the MIT influenced by metal-ion doping.
In $\mathrm{V}_{1-x}\mathrm{M}_x\mathrm{O}_2$, $N_d-N_a$ in Eq.~(\ref{eq:renormT0doped}) can be roughly approximated to be $x(v_d - v_{\mathrm{V}^{4+}})$, that is, all the dopants are ionized which corresponds to impurity levels very close to the bottom of the conduction band or the top of the valence band.

We first acquire the values of $\Theta_1$ and $\Theta_2$ by fitting the simulation results to the experimentally measured $T$-$x$ phase diagrams of $\mathrm{V}_{1-x}\mathrm{Ti}_x\mathrm{O}_2$~\cite{Wu15Decoupling} and $\mathrm{V}_{1-x}\mathrm{Mo}_x\mathrm{O}_2$~\cite{Patridge12Elucidating,Whittaker11Microscopic}.
The radius of $\mathrm{V}^{4+}$ is $R_{\mathrm{V}^{4+}}=0.58~\mathrm{\AA}$~\cite{Shannon76Revised}.
The radius and the valence of $\mathrm{Ti}^{4+}$ are $R_d=0.605~\mathrm{\AA}$~\cite{Shannon76Revised} and $v_d=4$, respectively.
Thus $F_2=0$ and $\Theta_2$ is irrelevant for $\mathrm{V}_{1-x}\mathrm{Ti}_x\mathrm{O}_2$.
The fitted $T$-$x$ phase diagram of $\mathrm{V}_{1-x}\mathrm{Ti}_x\mathrm{O}_2$ is shown in Fig.~\ref{fig:TiMo}(a), yielding $\Theta_1 = 16 k_B T_c$.
The experiment showed that the transition temperature reaches a minimum at $x\sim 2.8\%$ and then increases as $x$ increases~\cite{Wu15Decoupling}.
We did not fit the calculation result to the experimental measurement at high doping concentrations ($x \geq 2.8\%$), because essentially the theory is only valid for dilute doping.
Based on this, $\Theta_2$ can be fitted to the $T$-$x$ phase diagram of $\mathrm{V}_{1-x}\mathrm{Mo}_x\mathrm{O}_2$.
The X-ray absorption near-edge structure and extended X-ray absorption fine structure measurement determined that Mo in $\mathrm{V}_{1-x}\mathrm{Mo}_x\mathrm{O}_2$ is pentavalent~\cite{Patridge12Elucidating}, i.e., $v_d = 5$.
The radius of the $\mathrm{Mo}^{5+}$ ion is $R_d = 0.61~\mathrm{\AA}$~\cite{Shannon76Revised}.
The fitting result is shown in Fig.~\ref{fig:TiMo}(b), and the fitted $\Theta_2=20 k_BT_c$.

Using the fitted $\Theta_1$ and $\Theta_2$, we calculate the $T$-$x$ phase diagrams of $\mathrm{V}_{1-x}\mathrm{W}_x\mathrm{O}_2$ and $\mathrm{V}_{1-x}\mathrm{Cr}_x\mathrm{O}_2$.
The radius and the valence of $\mathrm{W}^{6+}$ are $R_d=0.60~\mathrm{\AA}$~\cite{Shannon76Revised} and $v_d=6$, respectively.
With these data, the calculated $T$-$x$ phase diagram of $\mathrm{V}_{1-x}\mathrm{W}_x\mathrm{O}_2$ is presented in Fig.~\ref{fig:WCr}(a), showing fair agreement with the experiments at low doping concentrations~\cite{Tan12Unraveling,Shi07Preparation,Horlin72Electrical,Shibuya10Metal,Wu11Temperature,Whittaker11Distinctive}.
It is not surprising that the calculation result does not agree well with the experiments at high doping concentrations ($x \gtrsim 2\%$), since the theory only addresses the case of dilute doping.
Only the R and the M1 phases appear on the phase diagram.
The calculated MIT temperature decreases nearly linearly at a large rate $-67~\mathrm{K/a.t.\%}$ as the W concentration increases.

Cr in $\mathrm{V}_{1-x}\mathrm{Cr}_x\mathrm{O}_2$ is trivalent, i.e., $v_d=3$.
The radius of the $\mathrm{Cr}^{3+}$ is $R_d = 0.615~\mathrm{\AA}$~\cite{Shannon76Revised}.
Using these data, we calculate the $T$-$x$ phase diagram of $\mathrm{V}_{1-x}\mathrm{Cr}_x\mathrm{O}_2$ and the result is shown in Fig.~\ref{fig:WCr}(b).
The experiment identified a new monoclinic (M3) phase between the M1 and the M2 phases~\cite{Marezio72Structural}.
The M2 and the M3 phases are separated by a discontinuity in volume but with no change in symmetry~\cite{Marezio72Structural}.
The calculation formally identifies the R, M1, and M2 phases; the M2 and the M3 phases cannot be distinguished by their order parameters within this phase-field model since they have the same symmetry.
The calculated phase diagram is in reasonable agreement with the experiment~\cite{Marezio72Structural}.
Some other experiments found that the discontinuity of the M1--M2 phase transition is reduced by the presence of transitional T phase between the M1 and the M2 phases on the phase diagram~\cite{Villeneuve73Contribution,Pouget76Metal}, however the T phase may be metastable compared to the M1 and the M2 phases~\cite{Park13Measurement}.
The calculated MIT temperature (the R--M2 transition temperature) increases as the Cr concentration increases, consistent with the experiment.

In Fig.~\ref{fig:WCr}, we successively reproduced the experimental observations that the trivalent dopant induces the intermediate M2 phase in addition to the R and the M1 phases while the pentavalent dopant does not.
In the phase-field model, this is due to the difference between $|v_d-v_{\mathrm{V}^{4+}}|$ appearing in Eq.~(\ref{eq:renormT0doped}) and $v_d-v_{\mathrm{V}^{4+}}$ appearing in Eq.~(\ref{eq:renormT0struct}).
For $|v_d-v_{\mathrm{V}^{4+}}|\sim 1$, $\Theta_1$ term is negligible compared to $\Theta_2$ term in Eq.~(\ref{eq:renormT0struct}).
If $v_d > v_{\mathrm{V}^{4+}}$ which is the case for pentavalent and hexavalent dopants, $T_0$ and $T_0'$ are both renormalized down.
This leads to a simple downshift of the transition temperature and thus no M2 phase appears.
If $v_d < v_{\mathrm{V}^{4+}}$ which is the case for trivalent dopants, $T_0$ and $T_0'$ are renormalized up and down, respectively.
The structural and the electronic instabilities are separated further in temperature, resulting in the appearance of the intermediate M2 phase (and possibly other intermediate M3 and T phases) between the two instabilities.

\section{Conclusion}
We formulated a Landau potential addressing the doping-induced regulation of the MIT in $\mathrm{VO}_2$.
The effect of the charge doping on the electron correlation is accounted for naturally by the addition of the free energy of free carriers into the total free energy.
The effects of doping on the lattice structure are abstracted as a combination of the volume effect and the Peierls effect, which are described by two coupling energies with the coupling constant fitted to experimentally measured phase diagrams of $\mathrm{V}_{1-x}\mathrm{Ti}_x\mathrm{O}_2$ and $\mathrm{V}_{1-x}\mathrm{W}_x\mathrm{O}_2$.
The Landau potential yields the $T$-$x$ phase diagrams of $\mathrm{V}_{1-x}\mathrm{Mo}_x\mathrm{O}_2$ and $\mathrm{V}_{1-x}\mathrm{Cr}_x\mathrm{O}_2$ consistent with the experiments.
The dramatic reduction of the transition temperature induced by the pentavalent and hexavalent  ion doping is caused by the simultaneous suppression of the stability of the strongly correlated electrons and V--V dimerization, while the emergence of intermediate phases induced by the trivalent ion doping is related to the separation in temperature of the electronic and structural instabilities.

\begin{acknowledgments}
This work was funded by the Penn State MRSEC, Center for Nanoscale Science, under the award NSF DMR-1420620.
\end{acknowledgments}

\bibliography{VO2doping_ref}

\end{document}